\documentclass[12pt]{article}
\input epsf.sty
\usepackage{mathrsfs}
\usepackage{amsmath}
\usepackage{mathrsfs}
\usepackage{amssymb}
\usepackage{color}
\usepackage{psfrag}
\usepackage{graphicx}
\usepackage{subfigure}

\usepackage{rotating}

\newcommand{\bea}{\begin{eqnarray}}
\newcommand{\eea}{\end{eqnarray}}
\newcommand{\be}{\begin{equation}}
\newcommand{\ee}{\end{equation}}
\newcommand{\vs}[1]{\vspace{#1 mm}}

\renewcommand{\a}{\alpha}
\renewcommand{\b}{\beta}

\renewcommand{\d}{\delta}

\newcommand{\dsl}{\pa \kern-0.5em /}

\newcommand{\half}{\frac{1}{2}}
\newcommand{\pa}{\partial}

\newcommand{\nn}{\nonumber\\}

\begin{document}
\topmargin 0pt
\oddsidemargin 0mm

\begin{flushright}


\end{flushright}

\vspace{2mm}

\begin{center}
{\Large \bf Interpolating solution from AdS$_5$ to hyperscaling violating Lifshitz space-time}  

\vs{10}

{Parijat Dey\footnote{E-mail: parijat.dey@saha.ac.in} and 
Shibaji Roy\footnote{E-mail: shibaji.roy@saha.ac.in}}

 \vspace{4mm}

{\em

 Saha Institute of Nuclear Physics,
 1/AF Bidhannagar, Calcutta-700 064, India\\}

\end{center}

\vs{10}

\begin{abstract}        
We construct two interpolating solutions in type II string theory which interpolate between
an AdS$_5$ in the UV and a hyperscaling violating three (spatial) dimensional Lifshitz space-time in the IR. 
The first solution is non-supersymmetric and is obtained from a known intersecting non-supersymmetric
D3-brane with chargeless D0-brane solution of type IIB string theory, by restricting some parameters
characterizing the solution and going to a new coordinate. In the IR the dilaton is non-constant
in general and the metric is three (spatial) dimensional hyperscaling violating Lifshitz with dynamical 
critical exponent $z=(3+3\gamma)/(3-\gamma)$ and hyperscaling violation exponent $\theta = 12/(3-\gamma)$, 
where $\gamma$ is a real parameter and can take continuous values from $-1$ to $+1$. At the two extreme
values, i.e., for $\gamma = \pm 1$, the dilaton is constant. On physical grounds it is found that $\gamma$
must be restricted to lie in the range $-1\leq \gamma \leq 0$. The second solution is supersymmetric and is
obtained from the known F-D2 bound state solution of type IIA string theory by zooming into a particular region 
of space. In the UV, the proper description is given in a T-dual frame which is AdS$_5$, whereas in the
IR, it gives a three dimensional hyperscaling violating Lifshitz with $z=3$ and $\theta=2=d-1$.

\end{abstract}

\newpage
\noindent {\it 1.  Introduction} : Conventional theoretical tools are inadequate to give explanations of the physics
behind the strongly
interacting quantum many body systems. Holography plays a crucial role in dealing with such systems by mapping
them to some gravitational theories in one dimension higher. This goes by the name AdS/CFT correspondence
\cite{Maldacena:1997re} or more generally gauge/gravity duality \cite{Aharony:1999ti} where a strongly coupled 
field theory in $D$ dimensions is dual to
a weakly coupled gravity theory in $D+1$ dimensions. This duality has broad applications not only for relativistic
systems like QCD \cite{CasalderreySolana:2011us}, but also for the  
non-relativistic strongly interacting condensed matter systems \cite{Hartnoll:2009sz} and can be used to understand 
the physics behind
such systems. A particular class of latter systems which has attracted a lot of interests in recent times are those
having Lifshitz and Lifshitz-like (with hyperscaling violation) scaling symmetries at their
quantum critical point. Properties of these systems can be understood by holographically constructing their gravity
duals which show the same scaling symmetry as an isometry (upto a conformal transformation).

In addition to space-time translation and spatial rotation, Lifshitz symmetry is a non-relativistic scaling symmetry 
where space and time scale differently $t \to \lambda^z t$ and $x^i \to \lambda x^i$ ($i=1,2,\ldots,d$, with $d$ being the 
spatial dimension of the theory, $\lambda$, the scaling parameter and $z (\neq 1)$, the dynamical critical exponent). For
$d=2$, the gravitational theory whose solution (metric) exhibits the above scaling symmetry as an isometry has been 
given in \cite{Kachru:2008yh,Taylor:2008tg}. Similar metrics have also been shown to arise from different solutions 
of string theory,
(gauged) supergravity in \cite{Hartnoll:2009ns}. When systems have Lifshitz-like scaling symmetry, the physical quantities
do not scale as in the case of Lifshitz symmetry, but they scale as if the dimension of the system is reduced by
a parameter called the hyperscaling violation parameter ($\theta$) \cite{Fisher:1986zz}. In the corresponding 
gravitational theory the solution
(or the metric) is not invariant under the scaling we mentioned, but changes upto a conformal factor 
$ds \to \lambda^{\theta/d}\, ds$ \cite{Charmousis:2010zz,Dong:2012se,Ogawa:2011bz,Huijse:2011ef}. These metrics 
have also been shown to arise from various string theory and (gauged) supergravity
solutions in \cite{Dey:2012tg,Narayan:2012hk}. A class of systems belonging to this group satisfying $\theta=d-1$ is 
particularly interesting as 
they give logarithmic violation of the area law of the entanglement entropy \cite{Ogawa:2011bz} which in turn indicates 
the presence of a hidden fermi surface in the system \cite{Huijse:2011ef}.

In the previous works \cite{Dey:2012tg} we have shown how space-time having Lifshitz-like symmetry arises from various 
supersymmetric string
theory solutions. In this paper, among other things, we will show how similar symmetry can arise from some non-supersymmetric 
string theory
solution\footnote{These solutions in both asymptotically flat as well as non-flat spaces were originally constructed in arbitrary 
dimensions in \cite{Zhou:1999nm} (see also \cite{Ivashchuk:2001ra}). Various properties of these solutions, including BPS limits, 
brane-antibrane interpretations 
and open and closed string tachyon condensations were studied in \cite{Brax:2000cf,Lu:2007bu}. Similar non-supersymmetric 
brane-like solutions in asymptotically
AdS$_5$ $\times$ S$^5$ spaces have been constructed in \cite{Russo:1998hx,Kehagias:1999tr,Myers:1999psa,Constable:1999ch}. However, 
no stability analysis has been performed for these 
non-supersymmetric solutions. The main reason could be that these solutions have naked singularities and might be called unphysical,
however, their status in full string theory is not at all clearly understood. For extended solutions of the type we are going to discuss 
and also
for asymptotically AdS solutions the dynamical stability \cite{Gregory:1993vy} is usually correlated with the thermodynamic stability 
\cite{Gubser:2000ec} of the 
solutions. As we will discuss later, even the thermodynamic stability is not helpful to conclude the stability of the solutions 
considered here.}.
However, Lifshitz and Lifshitz-like symmetries being non-relativistic symmetries we expect them to appear at low 
energy or at IR of the field theory and this implies by holography that it should appear when the inverse radial coordinate
$r \to \infty$ in the corresponding gravity theory. On the other hand, when we go to UV in the field theory, i.e.,
when we take $r \to 0$ in the gravity side we expect that the full relativistic symmetry will be restored. This in turn
implies that there must exist interpolating solutions which will interpolate between relativistic solution like AdS$_5$
in the UV ($r \to 0$) and some non-relativistic solution like Lifshitz or Lifshitz-like solution in the IR ($r \to \infty$). 
Indeed such
solutions have already been reported in the literature \cite{Azeyanagi:2009pr,Mateos:2011ix,Braviner:2011kz,Kumar:2012ui,Singh:2013iba,
Faedo:2013aoa,Kachru:2013voa,Ghodrati:2014spa}. In \cite{Azeyanagi:2009pr}, some non-supersymmetric string theory solution interpolating
between AdS$_5$ and an anisotropic Lifshitz has been constructed. In \cite{Braviner:2011kz}, various interpolating 
solutions of the type 
mentioned have been obtained in $D$ dimensional AdS gravity with massive vector field and also in ${\cal N}=4$, $D=6$ gauged
supergravity theory. More recently, a supersymmetric string theory solution interpolating between AdS$_5$ and a Lifshitz with
hyperscaling violation with $z=7$ and logarithmically running dilaton has been obtained \cite{Faedo:2013aoa}. 
Different non-supersymmetric solutions
interpolating between Bianchi attractor in the IR and AdS or Lifshitz in the UV have been constructed in \cite{Kachru:2013voa}.

In this paper we construct both a non-supersymmetric and a supersymmetric interpolating solutions in type II string theory 
which interpolate between AdS$_5$ in the UV and a Lifshitz with hyperscaling violation in the IR. For the non-supersymmetric case,
We start from a known non-supersymmetric
D3-brane intersecting with chargeless D0-brane solution of type IIB string theory \cite{Bai:2006vv,Lu:2007bu}. 
This solution is characterized by four
independent parameters. By going to a suitable coordinate, fixing one of the parameters and then zooming into a suitable region
we construct the solution which gives AdS$_5$ in the UV ($r \to 0$) and a one parameter family of Lifshitz with hyperscaling 
violation in the IR ($r \to \infty$). For the hyperscaling violating Lifshitz solution the dynamical critical exponent has the value
$z = (3+3\gamma)/(3-\gamma)$ and the hyperscaling violation exponent has the value $\theta = 12/(3-\gamma)$, where $\gamma$ is
a real parameter which takes continuous values in the range $-1 \leq \gamma \leq 1$. However, from the form of the energy momentum tensor 
of the boundary theory, we find that $\gamma$ must lie in the range $-1\leq \gamma \leq 0$, such that the energy and the pressure 
densities of the solution remain
positive semidefinite. The dynamical critical exponent, therefore,
lies in the range $0 \leq z \leq 1$ and the hyperscaling violation exponent lies in the range $3 \leq \theta \leq 4$. The spatial
dimension of the theory is $d=3$. The dilaton in general is not constant, but it becomes constant only at the two extremes when
$\gamma = \pm 1$. On the boundary the interpolating solution can be interpreted as a flow from ${\cal N}=4$, $D=4$ super 
Yang-Mills theory to a non-relativistic many body theory in three spatial dimensions having a hyperscaling violating Lifshitz 
symmetry.

For the supersymmetric case, on the other hand, we start from a known 1/4 supersymmetric F-D2 bound state solution 
of type IIA string theory. By zooming into a particular region of space for the F-D2 solution we construct the interpolating solution which 
interpolates between AdS$_5$ in the UV and a $d=3$ hyperscaling violating Lifshitz space-time in the IR with $z=3$ and $\theta=2$.
Actually in the UV, the interpolating solution of F-D2 gives AdS$_4$ $\times$ R $\times$ S$^5$, with one of the directions (representing
R) getting smaller in size and should be compactified. In that case the solution becomes nine dimensional and in string frame is
given by AdS$_4\, \times$ S$^5$, with non-zero dilaton. However, if we want to remain in ten dimensions, the above supergravity solution is   
not good as it can not be trusted when the radius of the circle corresponding to the compact direction becomes small compared
to the string length.  The proper supergravity description then would be given by its T-dual configuration which can be seen to be
precisely AdS$_5$ $\times$ S$^5$.  

\vspace{.5cm}

\noindent{\it 2. The interpolating solutions} : This section consists of three subsections. In the first subsection we give the construction
of interpolating solution from some non-supersymmetric D3-brane solution of type IIB string theory. In the second subsection we
try to interpret the interpolation as a flow in the boundary theory. In the third subsection
we describe the construction of interpolating solution from some supersymmetric F-D2 bound state solution of type IIA string theory.

\vspace{.5cm}

\noindent{\it 2.1 Interpolation from non-supersymmetric solution}: To construct the interpolating solution we begin with certain 
intersecting 
non-supersymmetric solution of type II string theory discussed in \cite{Bai:2006vv,Lu:2007bu}. In \cite{Lu:2007bu}, we 
have constructed non-supersymmetric
D$p$-brane intersecting with chargeless D1-brane and D0-brane solution (given in Eqs.(4) -- (6) of that paper) for the purpose 
of showing an interpolation between
black D$p$-brane and KK bubble of nothing and interpreting the interpolation as closed string tachyon condensation\footnote{We remark that
the interpolation from black hole to bubble in gravity solutions occurs when we change some of the parameters of the solution continuously by hand
from one set of values to another. So, the interpolation occurs through a continuous series of singular solutions. However, this interpolation can 
be interpreted as a transition from
black hole to bubble by closed string tachyon condensation only under certain conditions and not in general \cite{Lu:2007bu}. The 
solution we deal with
here does not interpolate between black hole and bubble and so it cannot be interpreted as a transition under any condition.}. We will
use not exactly this solution but a variant of it so that there is no D1-brane present. This is achieved by setting the coefficient
of $(dx^1)^2$ and that of $(dx^i)^2$ for $i=2,3,\ldots,p$ equal. That means we put $\delta_0 = \delta_2$ in that solution.
Furthermore, we consider $p=3$, i.e., it represents non-supersymmetric D3-brane intersecting with chargeless D0-brane of type IIB
string theory. The solution can be easily read off from Eqs.(4) -- (6) in \cite{Lu:2007bu} and can be written 
as\footnote{It is possible to
take a double Wick rotation of this solution by taking $\rho \to it$ and $t \to -ix^4$ to obtain a real, time dependent, 
space-like D3-brane
solution which is anisotropic in $x^4$-direction. In a special coordinate system this solution has recently been shown 
\cite{Roy:2014mba} to
give a four dimensional de Sitter space upto a conformal transformation upon compatification on some six dimensional product space.},
\bea\label{d3d0}
ds^2 &=& - F^{-\half}(\rho) 
\left(\frac{H(\rho)}{\tilde H(\rho)}\right)^{\frac{3\delta_1}{8} + \frac{3}{2}\delta_0} dt^2 + F^{-\half}(\rho) 
\left(\frac{H(\rho)}{\tilde H(\rho)}\right)^{-\frac{5\delta_1}{8} - \half \delta_0}     
\sum_{i=1}^3 (dx^i)^2\nn 
& & \qquad\qquad\qquad\qquad\qquad + F^{\half}(\rho) (H(\rho)\tilde H(\rho))^{\half}\left(\frac{H(\rho)}{\tilde H(\rho)}\right)^{\frac
{3\delta_1}{8}}\left(d\rho^2 + \rho^2 d\Omega_5^2\right)\nn
e^{2(\phi-\phi_0)} &=& \left(\frac{H(\rho)}{\tilde H(\rho)}\right)^{\frac{\delta_1}{2} - 6\delta_0}, \qquad F_5 = (1 + \ast) Q {\rm Vol}
(\Omega_5)       
\eea
where the various functions appearing in the above solution are defined as,
\bea\label{functions}
H(\rho) &=& 1 + \frac{\omega^4}{\rho^4}, \qquad \tilde H(\rho) = 1 - \frac{\omega^4}{\rho^4}\nn
F(\rho) &=& \left(\frac{H(\rho)}{\tilde H(\rho)}\right)^\alpha \cosh^2\tilde\theta 
- \left(\frac{\tilde H(\rho)}{H(\rho)}\right)^\beta \sinh^2\tilde\theta.
\eea
In \eqref{d3d0} the metric is given in the Einstein frame, $\phi$ is the dilaton field and $\phi_0$ is its asymptotic value. 
$F_5$ is the self-dual five form, 
where $Q$ is the charge parameter and $\ast$ denotes the Hodge dual.
The solution is characterized by seven parameters, namely, $\a,\,\b,\,\d_0,\,\d_1,\,\omega,\,\tilde\theta,$ and $Q$. 
The equations of motion give us three relations among them and they are
\bea\label{relations}
& & \a-\b = -\frac{3}{2}\d_1\nn
& & \half\d_1^2 + \half \a(\a+\frac{3}{2}\d_1) + \frac{3}{2}\d_0^2 = (1-3\d_0^2)\frac{5}{4}\nn
& & Q =  4 \omega^4 (\a+\b)\sinh2\tilde\theta.
\eea
Therefore, there are actually four independent parameters in the solution\footnote{Here one might think that the solution contains
too many parameters in contradiction with the Birkhoff's theorem. However, we point out that such uniqueness theorem applies for
regular manifolds and not for singular manifolds as is the case here. The physical interpretations of these parameters are not well-understood.
In one interpretation they can be related to number of D3 branes, number of anti-D3 branes, number of D0 branes and a tachyon parameter 
\cite{Bai:2006vv}. Note that since the solution has non-zero D3 brane charge, the number of D3 branes and number of anti-D3 branes are
different and their difference gives the total D3 brane charge. On the other hand the solution has no D0 brane charge, as this is a non-BPS
D0 brane. We remark that although this is a solution of type IIB string theory, it contains a D0 brane, where this D0 brane is non-BPS.}. 
Also note that the solution has a singularity
at $\rho = \omega$ and so, the physical region is given by $\rho > \omega$. We now make a change of coordinate by the 
following relation,
\be\label{coordchange}
\rho = {\bar r} \left(\frac{1+ \sqrt{g({\bar r})}}{2}\right)^{\half}, \qquad {\rm where,} 
\quad g({\bar r}) = 1 + \frac{4\omega^4}{{\bar r}^4}
\equiv 1 + \frac{r_0^4}{{\bar r}^4}
\ee
and by this the singularity at $\rho = \omega$ has now been shifted to $\bar r =0$.    
In terms of this new coordinate various functions given in \eqref{functions} take the forms,
\bea\label{functionsnew}
& & H(\rho) = 1 + \frac{\omega^4}{\rho^4} = \frac{2\sqrt{g({\bar r})}}{1+\sqrt{g({\bar r})}}\nn
& & \tilde H(\rho) = 1 - \frac{\omega^4}{\rho^4} = \frac{2}{1+\sqrt{g({\bar r})}}\nn
& & F(\rho) = \left(\frac{H(\rho)}{\tilde H(\rho)}\right)^\alpha \cosh^2\tilde\theta - \left(\frac{\tilde H(\rho)}
{H(\rho)}\right)^\beta \sinh^2\tilde\theta
= G({\bar r}) g({\bar r})^{\frac{\alpha}{2} - 1}
\eea
where,
\be\label{Gfunction}
G({\bar r}) = 1 + \frac{r_0^4 \cosh^2\tilde\theta}{\bar r^4}
\ee
Note that in writing the last expression for the function $F(\rho)$ in \eqref{functionsnew} we have used the relation 
$\alpha + \beta = 2$. This actually
fixes one of the independent parameters of the solution and we will be left with only three independent parameters. 
Now using \eqref{functionsnew}
in the solution \eqref{d3d0}, we can rewrite the metric and the dilaton as follows,
\bea\label{d3d0new}
& & ds^2 = - G(\bar r)^{-\half} 
g(\bar r)^{\frac{1}{4} + \frac{3\delta_1}{8} + \frac{3\delta_0}{4}} dt^2 + G(\bar r)^{-\half} 
g(\bar r)^{\frac{1}{4} -\frac{\delta_1}{8} - \frac{\delta_0}{4}}     
\sum_{i=1}^3 (dx^i)^2 + G(\bar r)^{\half} \left(\frac{d{\bar r}^2}{g(\bar r)} + \bar r^2 d\Omega_5^2\right)\nn
& & e^{2(\phi-\phi_0)} = g(\bar r)^{\frac{\delta_1}{4} - 3\delta_0},
\eea
In writing the metric in \eqref{d3d0new} we have used the condition $\a+\b=2$ and also the first parameter 
relation given in \eqref{relations},
$\a-\b=-(3/2)\d_1$. Combining these two relations we obtain $\a = 1-(3/4)\d_1$ and this is used to eliminate 
the parameter $\a$ in the above metric. Using
this in the second relation given in \eqref{relations} we obtain a simpler parameter relation as
\be\label{simplerelation}
\frac{7}{8} \d_1^2 + 21 \d_0^2 = 3
\ee             
Now let us zoom into the region ${\bar r}^4 \sim r_0^4 \ll r_0^4 \cosh^2\tilde\theta$ (note that this implies $\cosh^2\tilde\theta \gg 1$), 
then the harmonic function
$G(\bar r) = 1 + r_0^4 \cosh^2\tilde\theta/{\bar r}^4 \approx r_0^4\cosh^2\tilde\theta/{\bar r}^4 \equiv R^4/{\bar r}^4$ and 
the metric in \eqref{d3d0new}
would therefore reduce to,
\be\label{newmetric}
ds^2 =  \frac{\bar r^2}{R^2} 
\left(-g(\bar r)^{\frac{1}{4} + \frac{3\delta_1}{8} + \frac{3\delta_0}{4}} dt^2 +  
g(\bar r)^{\frac{1}{4} -\frac{\delta_1}{8} - \frac{\delta_0}{4}}     
\sum_{i=1}^3 (dx^i)^2\right) + \frac{R^2}{\bar r^2} \left(\frac{d{\bar r}^2}{g(\bar r)} + \bar r^2 d\Omega_5^2\right)
\ee
The zooming in here is actually equivalent to taking the parameter $\tilde\theta \to \infty$, keeping the other parameter $r_0$ finite. 
This affects the function
$G(\bar r)$ as given above but does not change the form of the function $g(\bar r)$. The resulting solution \eqref{newmetric}
is an exact solution of type IIB string theory very similar to the near horizon limit of black D3-brane. Asymptotically the solution
\eqref{newmetric} has the AdS$_5$ $\times$ S$^5$ structure, but the solution
has a curvature singularity at $\bar r =0$ for the generic values of the parameters $\d_0$ and $\d_1$. The singularity is hidden behind a regular
horizon when the parameters take values $\d_0=-1/7$ and $\d_1=-12/7$ and in that case the solution \eqref{newmetric} can be shown to take
precisely the form of AdS$_5$ black hole times S$^5$. One can even get a regular solution in the form of bubble of nothing for these values of
the parameters by taking a double Wick rotation $t \to ix^1$ and $x^1 \to -it$, by making $x^1$ compact with periodicity $\pi R^2/r_0$. But
for other values of the parameters, the singularity cannot be hidden by introducing a blackening factor, as that will not be a string theory
solution any more. The black solution is already included in \eqref{newmetric} for the particular values of $\d_0$ and $\d_1$ as we mentioned
and therefore cannot be blackened again. Now changing the coordinate once more by $r = R^2/{\bar r}$, the metric in \eqref{newmetric} and the 
dilaton take the form,
\bea\label{newmetricdilaton}
& & ds^2 =  \frac{R^2}{r^2} 
\left(-g(r)^{\frac{1}{4} + \frac{3\delta_1}{8} + \frac{3\delta_0}{4}} dt^2 +  
g(r)^{\frac{1}{4} -\frac{\delta_1}{8} - \frac{\delta_0}{4}}     
\sum_{i=1}^3 (dx^i)^2 + \frac{dr^2}{g(r)}\right) + R^2 d\Omega_5^2 \nn  
& & e^{2(\phi-\phi_0)} = g(r)^{\frac{\delta_1}{4} - 3\delta_0}, \qquad {\rm where,} \quad g(r) = 1 + \frac{r_0^4 r^4}{R^8}
\eea                      
The above solution represents the interpolating solution which, as we will show, interpolates between AdS$_5$ in the UV
($r \to 0$) and a one parameter family of hyperscaling violating Lifshitz in the IR ($r \to \infty$). The first part
is obvious from \eqref{newmetricdilaton}. When we take $r \to 0$, the function $g(r) = 1 + r_0^4 r^4/R^8 \to 1$, since
both $r_0$ and $R$ are finite. The solution \eqref{newmetricdilaton} then takes the form,
\bea\label{ads}
& & ds^2 = \frac{-dt^2 + \sum_{i=1}^3 (dx^i)^2 + dr^2}{r^2} + d\Omega_5^2 \nn  
& & e^{2(\phi-\phi_0)} = 1,
\eea  
where we have put both $R$ and $r_0$ to unity for simplicity. Note that there is no parameter dependence in 
this case although the solution \eqref{newmetricdilaton} depends on three independent parameters as we mentioned before.
Also note that since the radius of S$^5$ is constant it gets decoupled and the dilaton is constant. Thus we get AdS$_5$
in the UV from the solution \eqref{newmetricdilaton}.

Now in the IR, i.e., when we take $r \to \infty$, the metric and the dilaton in \eqref{newmetricdilaton} take the form,
\bea\label{hvlif}
& & ds^2 =  \frac{1}{r^4} 
\left(-r^{3+3\gamma} dt^2 +  r^{3-\gamma}     
\sum_{i=1}^3 (dx^i)^2 + \frac{dr^2}{r^2}\right) + d\Omega_5^2 \nn  
& & e^{2(\phi-\phi_0)} = r^{\delta_1 - 12\delta_0}
\eea
where $\gamma = (1/2)\d_1 + \d_0$ and again we have put $R$ and $r_0$ to unity for simplicity.   
To show that the above metric has Lifshitz-like scaling symmetry, we need to make another coordinate transformation given
by
\be\label{trans}
u^2 = r^{\gamma-3}
\ee
The metric and the dilaton would then be given as,
\bea\label{hvlif1}
& & ds^2 =  u^{\frac{8}{3-\gamma}} 
\left(-\frac{dt^2}{u^{\frac{2(3+3\gamma)}{3-\gamma}}} + \frac{     
\sum_{i=1}^3 (dx^i)^2}{u^2} + \frac{4}{(3-\gamma)^2}\frac{du^2}{u^2}\right) + d\Omega_5^2 \nn  
& & e^{2(\phi-\phi_0)} = u^{-\frac{2(\d_1-12\d_0)}{3-\gamma}}
\eea
It is clear that the part of the metric within bracket given in \eqref{hvlif1} is invariant under the scaling
$t \to \lambda^z t \equiv \lambda^{(3+3\gamma)/(3-\gamma)} t$, $x^i \to \lambda x^i$, for $i=1,2,3$ and $u \to \lambda u$.
But the whole metric is not invariant and under the scaling it changes as $ds \to \lambda^{\theta/d} ds \equiv 
\lambda^{12/[3(3-\gamma)]} ds$, where $d=3$ is the spatial dimension of the boundary theory (since S$^5$ gets decoupled). 
Thus the dynamical critical exponent $z$, the hyperscaling violation exponent $\theta$ and the spatial dimension $d$ 
of the theory are
\be\label{ztheta}
z = \frac{3+3\gamma}{3-\gamma}, \qquad \theta = \frac{12}{3-\gamma}, \qquad d = 3
\ee
Since the solution \eqref{hvlif1} depends on $\d_0$ and $\d_1$, and they are related by the relation \eqref{simplerelation}
we therefore get a one parameter family of hyperscaling violating Lifshitz space-time in the IR ($r \to \infty$). One can 
intuitively try to understand why we have a one parameter solution \eqref{hvlif1}. Actually, all the solutions we have either
in \eqref{newmetricdilaton} or in \eqref{hvlif1} are four parameter solutions as we mentioned. Out of the four we have set $r_0=R=1$
and therefore we should be left with two. Now, since \eqref{hvlif1} can be seen to come from dimensional reduction of ten dimensional 
type IIB on S$^5$, the two parameters must be associated with the two scalars, one coming from the radius of S$^5$ and the
other coming from the ten dimensional dilaton. However, we have fixed the radius of S$^5$ by putting $\alpha+\beta=2$, and so we will be left
with one parameter $\gamma$ only as is the case here in \eqref{hvlif1}.  
Now,
as $\d_0$ and $\d_1$ are real parameters, it is easy to see from \eqref{simplerelation} that they cannot take arbitrary values.
In fact, $\d_0$ must lie in the range $0\leq |\d_0| \leq \sqrt{1/7} \simeq 0.378$, and $\d_1$ will automatically fall in the range
$0 \leq |\d_1| \leq \sqrt{24/7} \simeq 1.852$. Now to find the range of $z$ and $\theta$, we must find the range of the parameter
$\gamma = (1/2)\d_1+\d_0$. Using the relation \eqref{simplerelation}, it can be checked that the maximum value $\gamma$ can take is
$+1$ and this occurs when $\d_0=1/7$ and $\d_1=12/7$, on the other hand the minimum value $\gamma$ can take is $-1$ and in that case
$\d_0=-1/7$ and $\d_1=-12/7$. It is therefore clear that $\gamma$ can take any real value in the range $-1 \leq \gamma \leq +1$.
However, as we will see in the next section that the energy momentum tensor of the boundary theory will further restrict the values of
$\gamma$ to lie in the range $-1 \leq \gamma \leq 0$, otherwise, the energy and the pressure densities of the states associated with
the solutions will become negative.  This
implies that $z$ and $\theta$ will lie in the range $0\leq z\leq 1$ and $3\leq \theta \leq 4$. Since we have $d=3$ in this case
we therefore have $\theta \geq d$. So, when $z>0$, such kind of gravity theories ($\theta > d$) have been argued in \cite{Dong:2012se} 
to be unstable even though the parameters can be shown to satisfy the null energy condition. This may not be completely unexpected. 
After all, the Lifshitz-like solution is obtained here
from non-supersymmetric, singular solution of string theory. However, we find that the instability argument discussed in \cite{Dong:2012se} 
either from thermodynamics or from entanglement entropy may not straightforwardly apply here. 
The reason is, the gravity solutions we have considered are singular and it is not clear how to study thermodynamics and argue about its 
thermodynamic instability\footnote{On the other hand, a simple scaling argument for the Lifshitz theory with
hyperscaling violation shows that the thermal entropy of such theory must scale as $S \sim T^{(d-\theta)/z}$ and so the specific heat
for $\theta> d$ must be negative indicating a thermal instability. We thank Sandip Trivedi for pointing this out to us.}. One might
think that since \eqref{hvlif1} is an exact solution of string theory one can cloak the singularity at $u=0$ behind a regular horizon
by putting a Schwarzschild-like factor $f(u)$ and $f^{-1}(u)$ in front of $dt^2$ term and $du^2$ term respectively and thus study the
thermodynamics of the system. As we have mentioned earlier, this way of hiding the singularity may be possible, but will not
be a solution of string theory. Actually all string theory solutions must be of the form \eqref{newmetric} for various allowed values of
$\d_0$ and $\d_1$, but it is clear that hyperscaling violating Lifshitz solutions with blackening factor are not contained in it. 
It is, therefore, not clear how to write a finite temperature version of the solution \eqref{hvlif1} and obtain the form of entropy
(as given in footnote by scaling argument) from there. Also for entanglement entropy, it might be possible to formally
obtain an entanglement entropy for this singular system, but it can be shown that it does not satisfy the strong subadditivity 
condition\footnote{This is also true for any thermodynamically unstable system as discussed in \cite{Bhattacharya:2012mi}.}. 
So, it is not clear whether such a quantity can be called an entanglement entropy or not. Also we note from the solution \eqref{hvlif1} 
that the dilaton is in general not constant and $e^{\phi}$ has a power law behavior. So, the dilaton changes as $\phi \to \phi - 
(\d_1-12\d_0)/(3-\gamma)\log\lambda$ under the scaling we mentioned before.
However, when $\gamma$ takes the two extreme values $\pm 1$, $\d_1$ and $\d_0$ take values $\pm 12/7$ and $\pm 1/7$ respectively,
the dilaton becomes a constant. 

Thus we have shown that the interpolating solution given in \eqref{newmetricdilaton} indeed interpolates between AdS$_5$ in the UV
($r \to 0$) to a one parameter family of hyperscaling violating Lifshitz in three spatial dimensions in the IR ($r \to \infty$) with
$\theta > d$. 
We point out that the original solution \eqref{d3d0} has four independent parameters as we mentioned before and we fixed one of them by using
$\a+\b=2$, while obtaining the interpolating solution. Out of the remaining three we fixed two more when we used $r_0=1$ and $R=1$
for simplicity. This is the reason we obtained a one parameter family
of hyperscaling violating Lifshitz in the IR. In the UV limit, on the other hand, there was no parameter left and we obtained an AdS$_5$
solution with unit radius.

\vspace{.5cm}

\noindent{\it 2.2 Interpretation in the boundary theory} :               
The interpolating solution in the bulk can be interpreted as a flow in the boundary gauge theory. The dependence of
the bulk field on the inverse radial coordinate $r$ is interpreted as the dependence on the energy parameter in the 
boundary theory. In particular,
the dilaton being non-constant (except for $\gamma = \pm 1$) implies the running of the 't Hooft coupling\footnote{Similar behavior
has been noted in some different non-supersymmetric type IIB solution in 
\cite{Kehagias:1999tr,Constable:1999ch}.} since the dilaton couples to the 
dimension four operator Tr$(F^2)$ in the gauge theory. The dilaton approaching asymptotically ($r \to 0$) to a constant value, 
$\phi-\phi_0 \sim r^4$, can
be interpreted as the flow of the gauge theory to a UV-stable fixed point \cite{Kehagias:1999tr,Constable:1999ch}. In this case, 
as we have seen, the geometry \eqref{newmetricdilaton} 
reduces to AdS$_5$ \eqref{ads} and the corresponding gauge theory at the fixed point is ${\cal N}=4$, $D=4$ super Yang-Mills theory.
The solution \eqref{newmetricdilaton} away from $r=0$ can be regarded as a deformation of AdS$_5$, by the addition of 
various normalizable modes
like the dilaton and the various components of the metric. Note that although these additions preserve the original 
SO(6) invariance, they do not
preserve the SO(1,3) Lorentz invariance of the metric (since the coefficient of $g_{tt}$ is 
different from those of $g_{ii}$'s, where $i=1,2,3$). 
They also break the conformal invariance as well as supersymmetry \cite{Kehagias:1999tr,Constable:1999ch}. In terms of the 
boundary gauge theory, we can think of the additions of these 
terms as the deformation
of the ${\cal N}=4$, $D=4$ super Yang-Mills theory. The additions of various normalizable modes amounts to the 
turning on the expectation values
of various gauge invariant operators \cite{Balasubramanian:1998sn}. So, for example, the addition of dilaton gives 
an expectation value $\langle{\rm Tr}(F^2)\rangle \sim 
(\d_1-12\d_0)$ in terms of the dimensionless parameters of the solution, where $\d_1$ and $\d_0$ are related by 
the relation \eqref{simplerelation}.
Note that the expectation value is dimensionless and that is because we have set the dimensionful parameters 
$r_0$ and $R$ to unity. Also, the 
addition of the metric components give expectation values of an energy-momentum tensor \cite{Myers:1999psa} in the form 
$\langle T_{\mu\nu}\rangle \sim {\rm diag}
(-3\gamma,\,-\gamma,\,-\gamma,\,-\gamma)$, where $\mu,\nu = 0,1,2,3$. This energy-momentum tensor\footnote{Similar form of energy
momentum tensors are given in \cite{Myers:1999psa,Constable:1999ch}. Actually the solution we have discussed in section 2.1 is an 
isotropic version of the solution discussed in \cite{Constable:1999ch}. In fact they can be mapped to each other by a coordinate
transformation when the solution of \cite{Constable:1999ch} is made isotropic. Also we have fixed the value of the volume scalar of
the 5-sphere by a special choice of some of our parameters just for simplification but it is not necessary.} can be regarded as
arising due to the excitations above the ground state representing zero temperature ${\cal N}=4$, $D=4$ super Yang-Mills theory. 
Now since the sign of the energy momentum tensor depends on the sign of $\gamma$, it is clear that the states with $\gamma > 0$
are unphysical since that will give negative energy and pressure densities\footnote{However, we mention that the negative energy 
solutions can be obtained from \eqref{newmetricdilaton}
if we make a Wick rotation $t \to ix^1$ and $x^1 \to -it$ and compactify $x^1$. This way we get new solutions which breaks SO(3) 
symmetry down to SO(2) and the 
expectation value of the energy momentum tensor for these new solutions would be given as $\langle T_{\mu\nu}\rangle \sim 
{\rm diag}(+\gamma,\,+3\gamma,\,-\gamma,\,-\gamma)$. Now since $\gamma \leq 0$ are physical we see that the new solutions (with
$\gamma < 0$) are
negative energy solutions. $\gamma=-1$ is special since in this case the solution can be made regular if $x^1$ is made
compact with periodicity $\pi R^2/r_0$ as we mentioned earlier. This solution is called AdS `bubble of nothing'
or AdS soliton. In the boundary theory this negative energy manifests itself in the form of Casimir energy as shown in 
\cite{Myers:1999psa}.}. From the form of the energy momentum tensor given above it is clear that it 
breaks the Lorentz 
invariance of the theory and as the boundary theory flows to IR ($r \to \infty$), we end up with a fixed point having a 
non-relativistic symmetry of the form of Lifshitz 
scaling symmetry with hyperscaling violation. Note that the above energy-momentum tensor is actually 
traceless, i.e., $\eta^{\mu\nu}\langle T_{\mu\nu}\rangle
=0$ as it should be. From the form of energy-momentum tensor it is clear that when $\gamma=0$, 
it vanishes, indicative of the restoration of 
Lorentz or relativistic symmetry \cite{Constable:1999ch}. Indeed we find from \eqref{ztheta} that when 
$\gamma=0$, the dynamical critical exponent $z=1$ and we recover
relativistic symmetry.  This happens when $\d_1 = \pm 2\sqrt{6}/7$ and $\d_0 = \mp \sqrt{6}/7$. Note that 
when $\gamma=0$, energy density vanishes which means that this is a zero energy
state. Moreover, since this state has non-zero VEV of Tr$(F^2)$, supersymmetry is compeletely broken. So, this is 
very much like a QCD state at zero temperature. In fact, when $\gamma=0$, the solution \eqref{newmetricdilaton} can
be mapped exactly to eq.(2.1), (2.2) of \cite{Constable:1999ch} (for $\delta=2$ there) and eq.(3.10), (3.11) of 
\cite{Csaki:2006ji}. The boundary
theory with various of its properties, resembling those of QCD, have been discussed at length in both these references. We
would like to point out that in our study of flows starting from ${\cal N}=4$, $D=4$ super Yang-Mills theory, the QCD-like
theory for $\gamma=0$ appears as an intermediate state in the IR and in the deep IR we get a theory which is conformal to
AdS$_5$ $\times$ S$^5$. 
Also, we have 
seen that when $\gamma = - 1$, the dilaton
remains constant and in those cases we do not turn on $\langle{\rm Tr}(F^2)\rangle$ in the gauge theory. 
From the metric in 
\eqref{newmetricdilaton} it can be checked that when $\gamma = -1$, i.e., when $\d_1=-12/7$ and $\d_0=-1/7$, 
the metric reduces precisely to the AdS$_5$ black hole. The boundary
theory then represents ${\cal N}=4$, $D=4$ super Yang-Mills theory at finite temperature. For the generic values of 
$\gamma$, the solution given in \eqref{newmetricdilaton} can be mapped to the solution discussed in \cite{Kim:2007qk}.  The 
boundary theory in this case has been interpreted as a finite temperature QCD like theory, some whose properties have
been discussed there. Again, we point out that this theory appears only in the intermediate stage and in the deep IR
we get Lifshitz theories with hyperscaling violation as discussed in previous section.

We, therefore, conclude that the interpolating solution described in section 2.1 can be interpreted as a flow
in the boundary theory for all values of $\gamma$ in the range $-1 \leq \gamma \leq 0$. Starting from ${\cal N}=4$, 
$D=4$ super Yang-Mills theory in the UV, we end up with a one-parameter
(given by $\gamma$) family of non-supersymmetric Lifshitz theory with hyperscaling violation in the IR (this happens only for
$-1 < \gamma < 0$), where the various
exponents are given in \eqref{ztheta}. For $\gamma=0$, the IR theory still remains relativistic although the supersymmetry
is broken by the non-zero VEV of Tr$(F^2)$. In fact the gravity dual in this case can be seen from \eqref{hvlif1} to take
the form of conformal AdS$_5$ times S$^5$. On the other hand, for $\gamma=-1$, the theory flows to a finite temperature
Yang-Mills theory. Here again the supersymmetry is broken but we have $\langle{\rm Tr}(F^2)\rangle = 0$ and $\langle T_{\mu\nu}
\rangle \neq 0$. For all other values of $\gamma$, we have a flow to hyperscaling violating Lifshitz with $\theta>d$ in 
the IR. Like in the previous cases we always have broken supersymmetry and the flows are triggered by non-zero VEV
of Tr$(F^2)$ and the energy momentum tensor. Even though the final state is hyperscaling violating
Lifshitz with $\theta>d$, but as we have pointed out that this criterion may not apply in the present context of 
non-supersymmetric solutions of string theory to conclude that the gravity solution is unstable invalidating the existence 
of the flow in the boundary theory.    
\vspace{.5cm}

\noindent{\it 2.3 Interpolation from supersymmetric solution} : In subsection 2.1 we obtained a solution
which interterpolates between AdS$_5$ in the UV and a one parameter family of hyperscaling violating $d=3$
Lifshitz solution in the IR from certain non-supersymmetric solution of type IIB string theory. The Lifshitz
solution we obtained in that case has $\theta > d$ and might be unstable. In this subsection we obtain another
similar interpolating solution from some supersymmetric bound state solution (F-D2) of type IIA string theory and 
in this case the Lifshitz solution has $\theta = d-1$ as was noted before in \cite{Deyroy}.

The quarter supersymmetric intersecting F-D$p$ bound state solutions of type II string theory are given in
eq.(2.6) of ref.\cite{Deyroy} and were obtained from the standard D1-D5 solution of type IIB string theory
by applying a series of T- as well as S-duality symmetries. For the purpose of constructing the supersymmetric
interpolating solution we will take F-D2 solution of type IIA string theory which can be written from eq.(2.6)
of ref.\cite{Deyroy} for $p=2$ and has the following form,
\bea\label{fd2}
& & ds^2 = H_2^{\half}\left[-H_1^{-1}H_2^{-1} dt^2 + H_2^{-1} \sum_{i=1}^2 (dx^i)^2 + H_1^{-1} (dx^3)^2 + d\bar{r}^2 + \bar{r}^2
d\Omega_5^2\right]\nn
& & e^{2(\phi-\phi_0)} = \frac{H_2^{\half}}{H_1}, \quad B_{[2]} = \left(1-H_1^{-1}\right) dt \wedge dx^3, \quad A_{[3]} = 
\left(1-H_2^{-1}\right) dt \wedge dx^1 \wedge dx^2
\eea   
Here $H_{1,2}$ are two harmonic functions and are given as,
\be\label{har}
H_{1,2} = 1 + \frac{Q_{1,2}}{\bar{r}^4}
\ee
with $Q_{1,2}$ representing the charges associated with F-strings and D2-branes respectively. $B_{[2]}$ and $A_{[3]}$ are respectively
the NSNS 2-form and RR 3-form which couple to F-strings and D2-branes. It is clear from the solution above that F-strings lie
along $x^3$, whereas D2-branes lie along $x^1$ and $x^2$. Note that here F-strings are delocalized along $x^1$ and $x^2$ and D2-branes
are delocalized along $x^3$ and they together form a (3+1)-dimensional world-volume. Let us now zoom into the region,
\be\label{zoom}
\bar{r}^4 \sim Q_1 \ll Q_2.
\ee
Then, the harmonic function $H_2$ can be approximated as $H_2 \approx Q_2/\bar{r}^4$, whereas, $H_1$ remains the same. Therefore, the
metric and the other form-fields take the form,
\bea\label{fd2new}
& & ds^2 = \frac{Q_2^{\half}}{\bar{r}^2}\left[-\frac{\bar{r}^4}{Q_2}H_1^{-1} dt^2 +\frac{\bar{r}^4}{Q_2} \sum_{i=1}^2 (dx^i)^2 + 
H_1^{-1} (dx^3)^2 + d\bar{r}^2 + \bar{r}^2
d\Omega_5^2\right]\nn
& & e^{2(\phi-\phi_0)} = \frac{Q_2^{\half}}{\bar{r}^2 H_1}, \quad B_{[2]} = \left(1-H_1^{-1}\right) dt \wedge dx^3, \quad A_{[3]} = 
-\frac{\bar{r}^4}{Q_2}dt \wedge dx^1 \wedge dx^2
\eea         
Note that since the solution is no longer asymptotically flat we have dropped a constant in $A_{[3]}$. Now changing the coordinate
by $r = Q_2^{\half}/\bar{r}$, the solution \eqref{fd2new} reduces to,
\bea\label{fd2new1}
& & ds^2 = \frac{Q_2^{\half}}{r^2}\left[-H_1^{-1} dt^2 + \sum_{i=1}^2 (dx^i)^2 + dr^2\right] 
+ \frac{r^2}{Q_2^{\half}}H_1^{-1}(dx^3)^2 + Q_2^{\half} d\Omega_5^2\nn
& & e^{2(\phi-\phi_0)} = \frac{r^2}{Q_2^{\half}}H_1^{-1}, \quad B_{[2]} = \left(1-H_1^{-1}\right) dt \wedge dx^3, \quad A_{[3]} = 
-\frac{Q_2}{r^4} dt \wedge dx^1 \wedge dx^2
\eea
where $H_1 = 1 + (Q_1/Q_2^2)r^4$. Eq.\eqref{fd2new1} is the required interpolating solution\footnote{A T-dual version of this solution
has been given in \cite{Singh:2010cj} but the interpolation there is quite different from that given here.} and we will show how 
this interpolates
between AdS$_5$ in the UV ($r \to 0$) and a hyperscaling violating Lifshitz with $d=3$, $z=3$ and $\theta =2=d-1$ in the IR ($r \to
\infty$). It is clear from \eqref{fd2new1} that in the UV, as $r \to 0$, $H_1 \to 1$ and therefore the solution reduces to
\bea\label{fd2new2}
& & ds^2 = \frac{Q_2^{\half}}{r^2}\left[-dt^2 + \sum_{i=1}^2 (dx^i)^2 + dr^2\right] 
+ \frac{r^2}{Q_2^{\half}}(dx^3)^2 + Q_2^{\half} d\Omega_5^2\nn
& & e^{2(\phi-\phi_0)} = \frac{r^2}{Q_2^{\half}}, \quad B_{[2]} = 0, \quad A_{[3]} = 
-\frac{Q_2}{r^4} dt \wedge dx^1 \wedge dx^2
\eea
The metric has the structure AdS$_4\,\times$ R $\times$ S$^5$, where, R corresponds to $x^3$ direction and becomes small
as $ r \to 0$. In that case we can compactify $x^3$ direction and end up with a nine dimensional solution where the metric
will have AdS$_4\,\times$ S$^5$ structure (given above without $(dx^3)^2$ term) and the nine dimensional dilaton will have 
the form $e^{2(\hat{\phi}-\hat{\phi}_0)} = r/Q_2^{1/4}$ with the 3-form remaining the same. However, if we want to remain in ten 
dimensions, the above supergravity description is not good since in the UV, the radius along the compact $x^3$ direction
can become of the order of string length and the stringy effects will be important. The good description would be given in 
the T-dual frame, i.e., if we take T-duality along $x^3$ direction. It can be checked by taking T-duality along $x^3$ that,
the metric in \eqref{fd2new2} reduces precisely to AdS$_5\,\times$ S$^5$ and the dilaton becomes constant. The 3-form
becomes a 4-form, whose field-strength must be made self-dual. This is exactly the near horizon structure of large number
of coincident D3-brane configuration. This way in the UV we recover the AdS$_5$ structure. 

On the other hand, in the IR when $r \to \infty$, the harmonic function $H_1 \approx (Q_1/Q_2^2) r^4$ and therefore, the interpolating
solution \eqref{fd2new1} reduces to
\bea\label{fd2newir}
& & ds^2 = Q_2^{\frac{3}{2}}\left[-\frac{Q_2}{Q_1}\frac{dt^2}{r^6} + \frac{\sum_{i=1}^2 (dx^i)^2 + dr^2}{Q_2 r^2} + \frac{(dx^3)^2}{Q_1 r^2}\right] 
+ Q_2^{\half} d\Omega_5^2\nn
& & e^{2(\phi-\phi_0)} = \frac{Q_2^{\frac{3}{2}}}{Q_1 r^2}, \quad B_{[2]} = -\frac{Q_2^2}{Q_1 r^4} dt \wedge dx^3, \quad A_{[3]} = 
-\frac{Q_2}{r^4} dt \wedge dx^1 \wedge dx^2
\eea        
This configuration was obtained in \cite{Deyroy} in eq.(2.11) with $p=2$. However, note that to match exactly with the solution given
in \cite{Deyroy}, we have to replace $r$ by $Q_2^{\half} r$ because of slightly different notation we choose here. In order to see the
scaling symmetry we dimensionally reduce the metric to five dimensions and express the resulting metric in Einstein frame. It has the form,
\be\label{fd2in5}
ds^2 =  Q_1^{\frac{2}{3}}Q_2^{\frac{5}{3}} r^{\frac{4}{3}}\left[-\frac{Q_2}{Q_1}\frac{dt^2}{r^6} + \frac{\sum_{i=1}^2 (dx^i)^2 + dr^2}{Q_2 r^2} 
+ \frac{(dx^3)^2}{Q_1 r^2}\right]
\ee
Now from \eqref{fd2in5} we observe that under the scaling $t \to \lambda^3 t$, $x^{1,2,3} \to \lambda x^{1,2,3}$ and $r \to \lambda r$, the
part of the metric in the bracket is invariant, but the full metric is not. In fact, under the above scaling the metric changes as
$ds \to \lambda^{2/3} ds \equiv \lambda^{\theta/d} ds$, where $\theta$ is the hyperscaling violation exponent. We thus conclude that in the
IR, the interpolating solution gives a hyperscaling violating Lifshitz space-time with spatial dimension $d=3$, dynamical exponent $z=3$
and hyerscaling violating exponent $\theta = 2 = d-1$. 

It is, therefore, clear that the solution \eqref{fd2new1} indeed interpolates between AdS$_5$ in the UV and a hyperscaling violating Lifshitz
with $z=3$, $d=3$ and $\theta=2$ in the IR. Note that although the interpolating solution \eqref{fd2new1} belongs to type IIA string theory,
in the UV its gravity description is given by AdS$_5\,\times$ S$^5$ which is a solution of type IIB string theory. This is because the metric
in \eqref{fd2new1} does not have a good gravity description in the UV if we want to remain in ten space-time dimensions, and in that case
the only gravity description be given in the T-dual frame, as we have argued, which takes the solution to type IIB theory. But the problem
in this case is that because of the T-duality transformation, it is difficult to give the interpretation of this interpolation as a 
flow in the boundary theory, nevertheless, it will certainly be interesting to have a better understanding of this issue.

\vspace{.5cm}
 
\noindent{\it 3. Conclusion} : To summarize, in this paper we have constructed two type II string theory solutions which interpolate
between an AdS$_5$ metric in the UV and a hyperscaling violating Lifshitz metric in the IR. Our starting point in the first case was a known 
non-supersymmetric D3-brane intersecting with a chargeless D0-brane solution of type IIB string theory. This is a four parameter 
solution which gives the interpolating solution when we expressed it in a suitable coordinate and restricted one of the parameters
to a special value. The interpolating solution gives AdS$_5$ metric in the UV (when $r \to 0$) and a one parameter family of Lifshitz 
metric (in three spatial dimension) with hyperscaling violation in the IR (when $r \to \infty$). The transverse 5-dimensional space
which is a 5-sphere gets completely decoupled in the solution. The IR Lifshitz-like metric has a dynamical critical exponent $z = 
(3+3\gamma)/(3-\gamma)$ and a hyperscaling violation exponent $\theta = 12/(3-\gamma)$, where $\gamma$ is a real parameter and lies
in the range $-1 \leq \gamma \leq 1$. From physical arguments we have shown that $\gamma$ must be restricted to lie in the range
$-1 \leq \gamma \leq 0$ only. We, therefore, have $z$ lying in the range $0 \leq z \leq 1$ and $\theta$ lying in the range 
$3 \leq \theta \leq 4$. When $\gamma = - 1$, the dilaton is constant, otherwise dilaton is a function of inverse radial coordinate $r$.
When $\gamma=-1$, the interpolating solution reduces to the AdS$_5$ black hole. On the other hand, when $\gamma=0$, the Lorentz invariance
is restored and in that case the IR theory is not non-relativistic or Lifshitz-like, rather it is conformal to AdS$_5$, i.e., relativistic.
We have noted that since in the IR we have Lifshitz-like theory with $d=3$ and $3 \leq \theta \leq 4$, we have $\theta\geq d$. These
theories (for $\theta>d$ and $z>0$) are usually unstable \cite{Dong:2012se}. However, we have mentioned that the argument of 
Dong et. al. \cite{Dong:2012se}  
may not be as straightforward as we think. This is because the solution itself is singular and it is not clear how to study thermodynamics
and how to define entanglement entropy for the system with $\theta > d$.  
We have seen that the interpolation of the supergravity solution can be understood in the boundary theory 
as a flow, where on physical grounds we have excluded the positive values of $\gamma$. By this flow ${\cal N}=4$, $D=4$ 
super Yang-Mills theory in the ultraviolet goes over to an infrared fixed
point which is non-relativistic and given by the strongly interacting many body system with a Lifshitz-like (with hyperscaling violation) 
scaling symmetry. This flow is triggered by turning on the expectation values of the operator Tr$(F^2)$ and various components of an 
energy-momentum tensor $T_{\mu\nu}$. They break conformal invariance, supersymmetry and SO(1,3) Lorentz invariance of the UV theory.

In the second case we consider a quarter supersymmetric F-D2 bound state solution of type IIA string theory. By zooming into certain
region of space in this solution we have shown how to obtain an interpolating solution which interpolates between AdS$_5$ in the UV 
and a hyperscaling 
violating Lifshitz in $d=3$ with $z=3$ and $\theta=2$. Thus in the supersymmetric case, we indeed find a hyperscaling violating Lifshitz 
which has $\theta <d$, in fact, it has $\theta=d-1$ in contrast to the non-supersymmetric case where we found that the hyperscaling
violating Lifshitz has $\theta > d$. However, in this case the interpolation involves a T-duality transformation and that is the reason
it is difficult to interpret this interpolation in terms of certain RG flow in the boundary theory. It would be nice to have a better
understanding of this issue and we leave it for the future.   
\vspace{.5cm}

\noindent{\it Acknowledgements} : We would like to thank Sandip Trivedi for 
comments and a careful reading of an earlier version of this manuscript. We would also
like to thank Debashis Ghoshal and Harvendra Singh for discussions. One of the 
authors (PD) would like to acknowledge 
thankfully the financial support of the Council of Scientific and Industrial Research, India
(SPM-07/489 (0089)/2010-EMR-I).

\end{document}